\documentstyle[epsfig]{aipproc}

\begin{document}
\hspace*{3.6in}{\bf IUHET-380}\\
\hspace*{3.6in}{\bf January 1998}
\title{Precision Measurements of\\ 
Threshold Chargino Production\thanks{Presented
at the Workshop on Physics at the First Muon Collider and at the
Front End of a Muon Collider, November 6-9, 1997,
Fermi National Accelerator Laboratory.}}

\author{M. S. Berger}
\address{Indiana University\\
Bloomington Indiana 47405}

%\lefthead{LEFT head}
%\righthead{RIGHT head}
\maketitle

\begin{abstract}
We analyze the prospects at
a muon collider for measuring chargino masses in the 
$\mu^+\mu^-\to \tilde{\chi}^+\tilde{\chi}^-$ processes
in the threshold region. 
We find that a measurement of the lightest chargino mass to better than 
200~MeV is possible with 100~fb$^{-1}$ luminosity. 
The muon sneutrino mass 
can also be simultaneously measured to a few GeV.
\end{abstract}

\section*{Introduction}

Particle masses can be measured quite accurately by producing them near
threshold. This has been demonstrated recently at LEP II where $W$ pairs
were produced at $\sqrt{s}=161$~GeV, and a precise measurement of the $W$ 
mass has been obtained. We have recently shown that future
high-luminosity $\mu^+\mu^-$ colliders can measure the $W$ boson,
top quark and Higgs boson masses with high precision in the processes
$\ell^+\ell^-\to WW, t\overline{t}, ZH$\cite{wwthresh,zhthresh,ttbar}.
Threshold production of chargino pairs at a muon collider 
offers a possible way of measuring
the chargino mass and also the muon sneutrino mass that is involved in the 
production process\cite{prep}.

Muon colliders\cite{mupmumi,saus,montauk,sanfran,feas} 
could be especially useful tools in precision measurements of particle
masses, widths, and couplings.
Initial state radiation from muons is reduced compared to electrons, and 
muon colliders have negligible beamstrahlung.
The threshold regions for particle production depend on the particle
widths. In fact, one can in principle measure the widths of the $W$ boson, 
the top quark, and the 
Higgs boson width by performing the appropriate measurements of the 
production cross sections near threshold.
When the lightest chargino is dominantly gaugino, its width is 
usually negligibly small and the threshold cross section is 
controlled only by angular
momentum considerations and the characteristics of the colliding beam. 
The measurement
of the chargino mass via the threshold cross section has been considered 
previously for electron-positron machines in Ref.~\cite{oldthresh,nlcstudy}.
We consider the measurement at a muon collider with 
high luminosity, carefully taking into account the beam effects and 
reoptimizing cuts to eliminate the background in the threshold region. 
We assume here 
that the muon collider has a relatively modest beam energy spread of 
$R=0.1\%$, where $R$ is the rms spread of the energy of a muon 
beam\footnote{The most recent TESLA design
envisions a beam energy spread of $R=0.2\%$\cite{miller} while the
NLC design expects a beam energy spread of $R=1.0\%$. A high
energy $e^+e^-$ collider in the large VLHC tunnel would have a beam 
spread of $\sigma _E=0.26$~GeV\cite{norem} which should give numbers
precisions comparable to those considered here.}.
We assume that 100~fb$^{-1}$ integrated luminosity is available; high
luminosity is necessary if the threshold measurement is to prove
interesting.

A precision measurement of the chargino mass will be highly desirable to test
patterns of supersymmetry breaking. For example the relationship between
the lightest neutralino and the lightest chargino masses can be used to test
the existance of a universal soft SUSY-breking parameter.
The chargino pair production process has been investigated beyond the 
tree-level recently\cite{rc}. A precision measurement of the cross section
can test radiative corrections coming from heavy squarks.

\section*{Signal and Background}

A simultaneous measurement of the chargino and sneutrino masses requires a 
sampling of the cross section at at least two points. As in other threshold
measurements, the statistical precision on the chargino mass is maximized 
just above $2m_{\tilde{\chi}^{\pm}}$. However as is evident from Fig.~1, 
a change in the cross section at $\sqrt{s}=2m_{\tilde{\chi}^{\pm}}+1$~GeV
can be due to a variation in the sneutrino mass, so a second measurement of 
the cross section must be taken at a higher $\sqrt{s}$ where the dependence
of the cross section on the chargino mass and the slepton mass is different.
It turns out to be advantageous for the chargino mass measurement to choose 
this higher energy measurement at a point where the chargino cross section 
is not flat.

The precision that can be obtained in the chargino mass depends substantially
on the chargino mass itself: the heavier the chargino the smaller
the production cross section. The cross section also depends on the mass of the
sneutrino which appears in the $t$-channel. The contribution
from the sneutrino graph interferes destructively with the $s$-channel graphs.
If the 
lightest chargino is gaugino-dominated, then changing the parameters of the 
chargino mass matrix essentially changes the mass but not the chargino 
couplings significantly. Therefore one can envision a measurement of the 
cross section
that depends on just two parameters: the chargino mass $m_{\tilde{\chi}^{\pm}}$
and the sneutrino mass $m_{\tilde{\nu}}$.

The chargino decay mode is 
$\tilde{\chi}^{\pm}\to \tilde{\chi}^0f\overline{f}^\prime$.
If $m_{\tilde{\chi}^{\pm}}-m_{\tilde{\chi}^0}>M_W$ then $W$ exchange dominates
and the final state is comprised of 49\% purely hadronic events,
42\% mixed hadronic-leptonic events, and 9\% purely leptonic events (these
ratios are determined by the $W$ branching fractions). The width of the 
chargino has a  negligible impact on the threshold cross section 
even the two body decay is possible 
($m_{\tilde{\chi}^{\pm}}-m_{\tilde{\chi}^0}>M_W$) if the lightest chargino
is gaugino-dominated. There are several 
backgrounds to the chargino pair signal, the largest being 
$\mu^+\mu^-\to W^+W^-$. The cross section is reduced near threshold, so 
the cuts to reduce this background need to be reoptimized. 

One must worry about the level of the backgrounds, and the systematic error
that the residual background presents for the cross section measurement. 
Figure 1 indicates that with the amount of integrated luminosity we are 
assuming will be available for the measurement, the cross section is being 
measured to the few percent level, so an understanding of the background to 
at least this level is necessary.
The backgrounds to chargino pair-production have been investigated in 
Refs.~\cite{tfmyo,grivaz} where the signal efficiencies have been obtained 
for the various final states when the center-of-mass energy is 
$\sqrt{s}=500$~GeV. The primary background is $W$ pair production which is 
very large, but can be effectively eliminated because the $W$'s are produced 
in the very-forward direction. However, if the energy is reduced so that the
collider is operating in the chargino threshold region, then the effectiveness
of these cuts might be reduced (the signal events might be expected to be 
more spherical as well). Therefore the efficiencies should be reinvestigated
for the threshold measurement.
The overall 
signal efficiency of our cuts in Ref.~\cite{prep} 
is about 10\% for the fully hadronic 
decays.

We have assumed here that the chargino is lighter than the muon sneutrino.
If that is not the case, the chargino has a new decay mode:
$\tilde{\chi}^{\pm}\to \ell^{\pm}\tilde{\nu}$. The efficiency 
of the cuts against background would need to be reconsidered if this mode
is kinematically allowed.

\begin{figure}[htb]
\leavevmode
\begin{center}
\epsfxsize=3.75in\hspace{0in}\epsffile{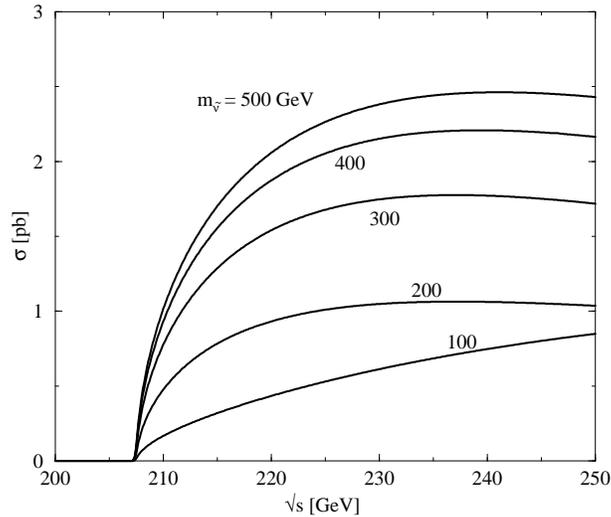}
\end{center}
\caption[]{\footnotesize\sf The threshold region of 
$\mu^+\mu^-\to \tilde{\chi}^+\tilde{\chi}^-$ for various sneutrino
masses, taking $M_2=100$~GeV and $\tan \beta =4$.  The rapid rise of
the cross section is due to the pair production of spin-1/2 particles
with small decay widths. The muon collider is assumed to have a beam
energy spread of $R=0.1\%$.}
\label{figure1}
\end{figure}

A further advantage of the threshold measurement is that the chargino mass
measurement is somewhat isolated from its subsequent decays. Distributions in
the final state observables, say e.g. $E_{jj}$ from the decay
$\tilde{\chi}^{\pm}\to \tilde{\chi}^0jj$\cite{tfmyo}, depend on the 
neutralino mass. The cross section for chargino pair production, on the 
other hand, is independent of the final state particles, and only the 
branching fractions and detector efficiencies for the various final states
impact this measurement (as indicated above,
if $m_{\tilde{\chi}^{\pm}}-m_{\tilde{\chi}^0}>M_W$
the branching fractions of chargino decay is given essentially in terms of 
the $W$ branching fractions).

The chargino production cross section decreases with increasing chargino mass.
Therefore the precision with which the mass can be measured is better at 
smaller values of the mass. Figure~2 shows the expect precision with 
100~fb$^{-1}$ integrated luminosity for sneutrino masses of 300 and 500~GeV. 
For a lighter sneutrino, for which the destructive interference between
the $s$-channel and $t$-channel graphs is more severe, the precision obtained
is reduced.
Furthermore inspection of Fig.~1 demonstrates the 
variability of the cross section is reduced for heavier sneutrino masses 
leading to a reduced precision measurement.
The sneutrino mass can be measured to about 6~GeV accuracy for 
$m_{\tilde{\nu}} = 300$~GeV and to about 20~GeV accuracy for 
$m_{\tilde{\nu}} = 500$~GeV. This provides an indirect method of 
measuring the sneutrino mass (the sneutrino 
might be too heavy to produce directly).

\begin{figure}[htb]
\leavevmode
\begin{center}
\epsfxsize=3.75in\hspace{0in}\epsffile{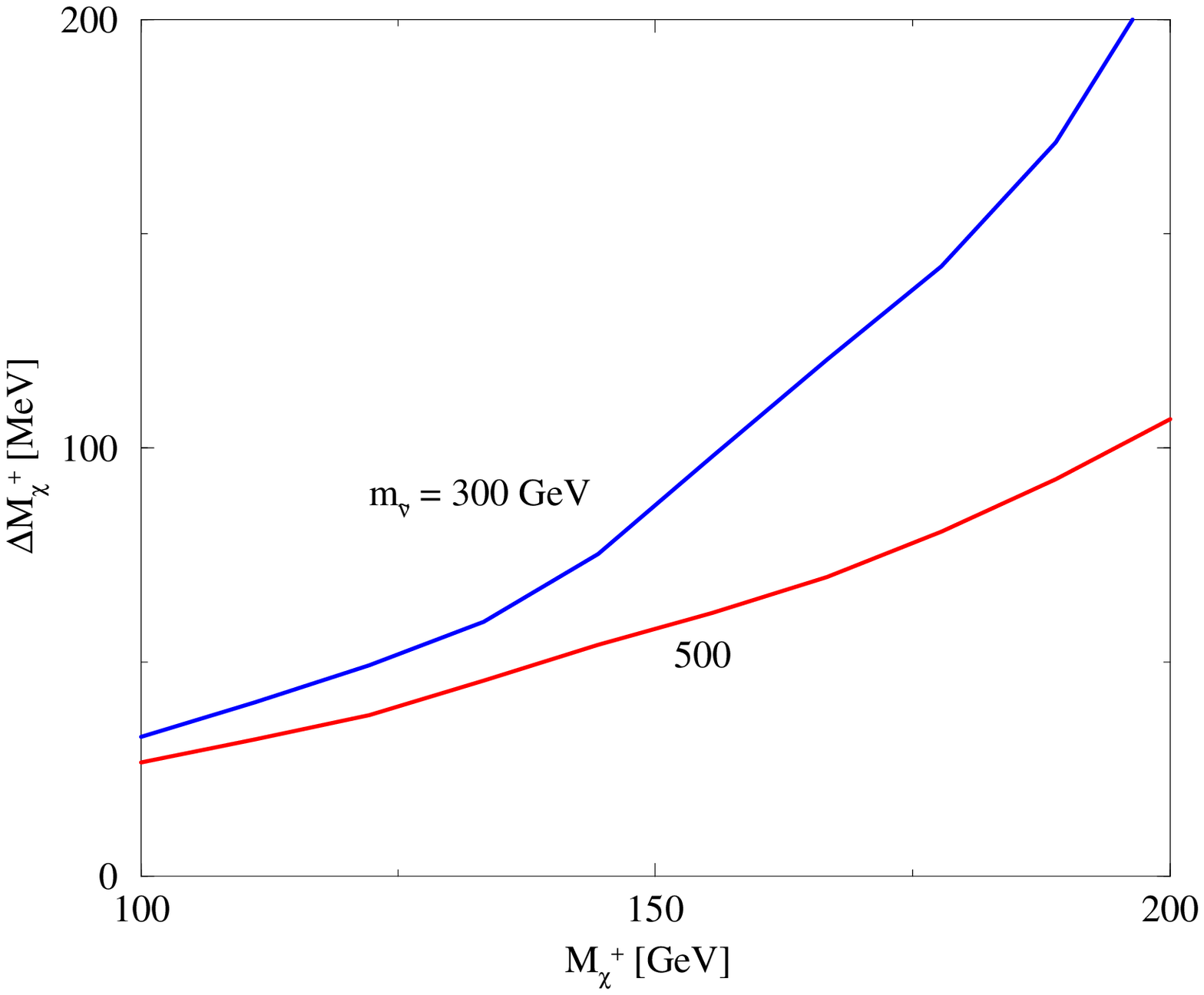}
\end{center}
\caption[]{\footnotesize\sf The $1\sigma$ precision obtainable in the 
chargino mass taking $m_{\tilde{\nu}} = 300$ and $500$~GeV. The precision is
better for {\it larger} sneutrino mass because the contribution from the
$t$-channel sneutrino exchange diagram destructively interferes with
the $s$-channel diagrams.}
\label{figure2}
\end{figure}

The result of a fit to the chargino cross section is shown in Fig.~3, taking
$M_2=120$~GeV and $\tan \beta =4$ and assuming an integrated 
luminosity of 100~fb$^{-1}$. 
The cross section is measured just above the threshold 
$\sqrt{s}=2m_{\tilde{\chi}^{\pm}}+1$~GeV, and at a point well
above the threshold, $\sqrt{s}=2m_{\tilde{\chi}^{\pm}}+20$~GeV.

\begin{figure}[htb]
\leavevmode
\begin{center}
\epsfxsize=3.75in\hspace{0in}\epsffile{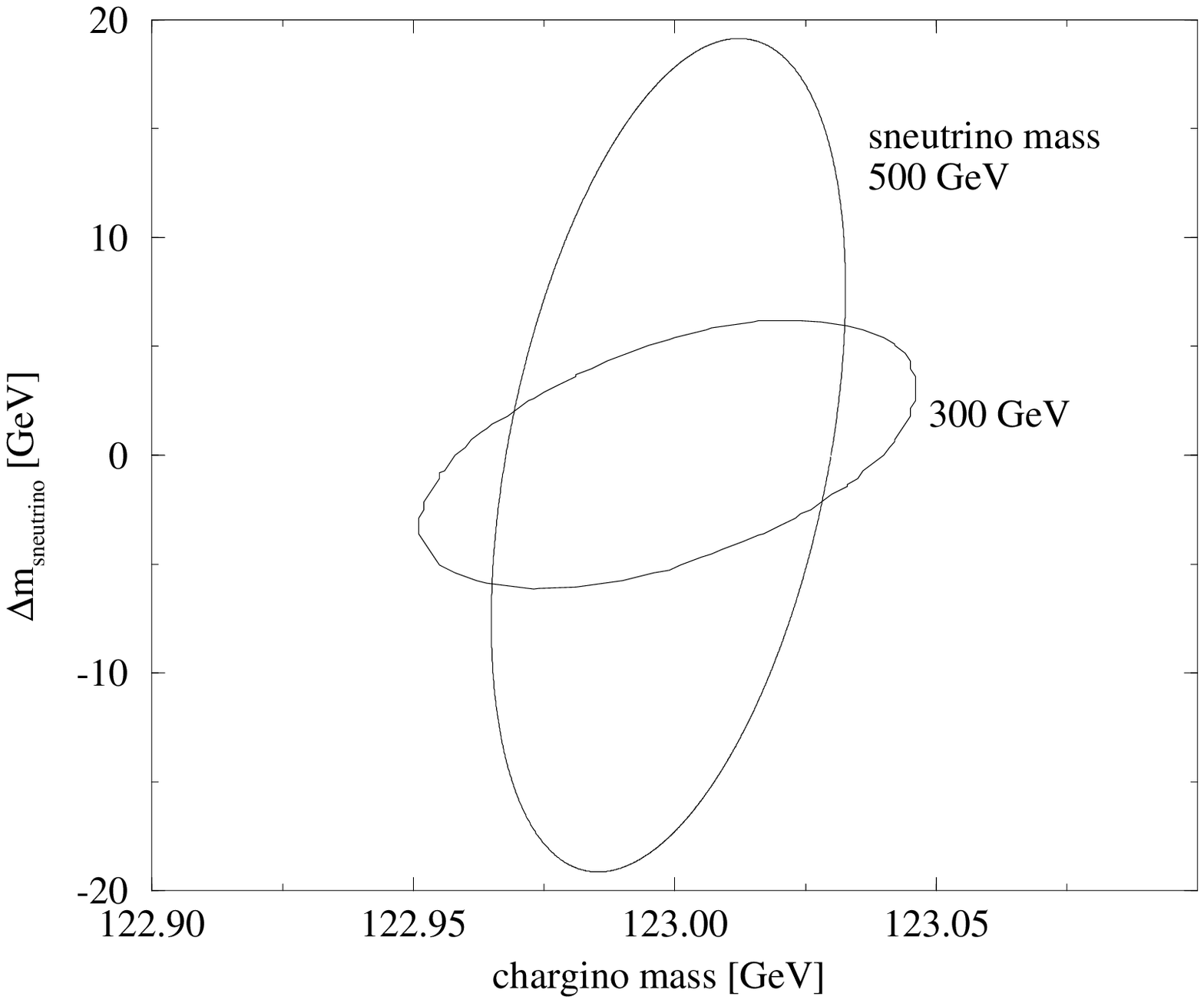}
\end{center}
\caption[]{\footnotesize\sf The $\Delta \chi^2 =1$ contours in the chargino 
mass - sneutrino mass plane, taking $M_2=120$~GeV, $\tan \beta =4$, and 
$m_{\tilde{\nu}}=300$ and $500$~GeV. This gives a chargino mass
$m_{\tilde{\chi}^{\pm}}\approx 123$~GeV.
The curves assume 50~fb$^{-1}$ 
of integrated
luminosity is devoted to $\sqrt{s}=2m_{\tilde{\chi}^{\pm}}+1$~GeV, and 
50~fb$^{-1}$ is applied at $\sqrt{s}=2m_{\tilde{\chi}^{\pm}}+20$~GeV.
The chargino mass determination is better for higher sneutrino mass since
the sneutrino exchange diagrams interferes destructively with the $s$-channel
diagrams.}
\label{figure3}
\end{figure}

\section*{Polarization}
It is expected that the both beams of a muon collider can be partially 
polarized, although with some loss of luminosity\cite{feas}. This could prove
a useful tool for measuring the gaugino and Higgsino components of the
chargino.
When the chargino is gaugino-dominated, it couples to the left-handed $\mu^-$
because the chargino is then dominantly the partner to the $W$. Since the $WW$ 
background can be reduced by having substantial right-handed polarization, 
some improvement can be expected in the chargino mass precision, especially in
the case where the chargino is higgsino-dominated.
Only the gaugino component of the chargino couples to the $t$-channel sneutrino
exchange, and this can be turned off by operating with polarized $\mu$
beams. 

For the gaugino-dominated chargino considered here, both the signal and 
background are approximately 
proportional to $(1-P)^2$ where $P$ is the polarization of 
the two muon beams. So for fully polarized $\mu ^+$ and $\mu^- $ beams
one can improve the mass determination by a factor of two.

\section*{Conclusions}
We have shown that a measurement of the lightest chargino mass to better than 
200~MeV is possible by measuring the pair production cross section near 
threshold at a muon collider with 100~fb$^{-1}$ luminosity. 
This is much better than other techniques. The muon sneutrino mass 
can also be simultaneously measured to a few GeV.

\section*{Acknowledgments}

I thank V.~Barger and T.~Han for a pleasant collaboration
on the issues reported here.
This work was supported in part by the U.S. Department of Energy
under Grant
No.~DE-FG02-91ER40661.


\begin{references}

\bibitem{wwthresh} V.~Barger, M.S.~Berger, J.F.~Gunion and T.~Han, 
Phys.\ Rev.\ {\bf D56}, 1714 (1997).

\bibitem{zhthresh} V.~Barger, M.S.~Berger, J.F.~Gunion and T.~Han, 
Phys.\ Rev.\ Lett.\ {\bf 78}, 3991 (1997).

\bibitem{ttbar}
M.S.~Berger, talk presented at the
{\it Workshop on Particle Theory and Phenomenology:
Physics of the Top Quark}, Iowa State University,
May 25--26, 1995, hep-ph/9508209.

\bibitem{prep} V.~Barger, M.S.~Berger and T.~Han, hep-ph/9801410.

\bibitem{mupmumi} {\it Proceedings of the First Workshop on the Physics
Potential
and Development of $\mu^+\mu^-$ Colliders}, Napa, California (1992), Nucl.\
Instru.\ and Meth.\ {\bf A350}, 24 (1994).

\bibitem{saus} {\it Proceedings of the Second Workshop on the Physics
Potential and Development of $\mu^+\mu^-$ Colliders}, Sausalito, California
(1994), ed.\ by D.~Cline, American Institute of Physics Conference
Proceedings 352.

\bibitem{montauk} {\it Proceedings of the 9th Advanced ICFA Beam Dynamics
Workshop: Beam Dynamics and Technology Issues for $\mu^+\mu^-$ Colliders},
Montauk, Long Island, (1995), to be bibitem.

\bibitem{sanfran} {\it Proceedings of the Symposium on Physics Potential and
Development of $\mu^+\mu^-$ Colliders}, San Francisco, California,
December 13-15, 1995.

\bibitem{feas} {\it $\mu^+\mu^-$ Collider: A Feasibility Study},
Snowmass, Colorado, July, 1996.

\bibitem{oldthresh} A.~Leike, Int.\ J.\ Mod.\ Phys.\ {\bf A3}, 2895 (1988).

\bibitem{nlcstudy} {\it Physics with $e^+e^-$ Linear Colliders},
by ECFA/DESY LC Physics Working Group (E. Accomando et al.), DESY-97-100, 
May 1997, hep-ph/9705442. 

\bibitem{miller} D.~Miller, private communication.

\bibitem{norem} J.~Norem, private communication and\\ 
http://www-ap.fnal.gov/VLHC/electrons/index.html.

\bibitem{rc} P.~Chankowski, Phys.\ Rev.\ {\bf D41}, 2877 (1990); 
M.~M.~Nojiri, K.~Fujii and T.~Tsukamoto, Phys.\ Rev.\ {\bf D54}, 6756 (1996);
H.-C.~Cheng, J.~L.~Feng and N.~Polonsky, hep-ph/9706476;  hep-ph/9706438;
M.~A.~Diaz, S.~F.~King and D.~A.~Ross, hep-ph/9711307.

\bibitem{tfmyo} T.~Tsukamoto, K.~Fujii, H.~Murayama, M.~Yamaguchi and 
Y.~Okada, Phys.\ Rev.\ {\bf D51}, 3153 (1995).

\bibitem{grivaz} J.-F.~Grivaz, preprint LAL 91-63, Talk at the Workshop on 
Physics and Experiments with Linear Colliders, Saariselka, Finland, 
9-14 September 1991.

\end{references}
\end{document}